\documentclass{ifacconf}

\usepackage{graphicx}      
\usepackage{natbib}        
\usepackage{amsmath}       
\usepackage{amssymb}       
\usepackage[utf8]{inputenc}

\usepackage{xspace}
\usepackage{soul}
\usepackage{mathtools}
\usepackage{makecell}
\usepackage{nicefrac}
\usepackage{float}
\usepackage{wrapfig}
\usepackage{color}

\newcommand{\U}{\ensuremath{\mathbb{U}}}

\let\emptyset\varnothing

\newtheorem{definition}{Definition}
\newtheorem{proposition}{Proposition}
\newtheorem{theorem}{Theorem}
\newtheorem{remark}{Remark}

\DeclareMathOperator*{\arginf}{arg\,inf}

\usepackage{textcomp}

\definecolor{dgreen}{rgb}{0.0, 0.5, 0.0}

\makeatletter
\newcommand{\subalign}[1]{%
	\vcenter{%
		\Let@ \restore@math@cr \default@tag
		\baselineskip\fontdimen10 \scriptfont\tw@
		\advance\baselineskip\fontdimen12 \scriptfont\tw@
		\lineskip\thr@@\fontdimen8 \scriptfont\thr@@
		\lineskiplimit\lineskip
		\ialign{\hfil$\m@th\scriptstyle##$&$\m@th\scriptstyle{}##$\crcr
			#1\crcr
		}%
	}
}

\usepackage{tcolorbox}

\newcommand{\DAM}{\left[\partial \mathcal{A}\right]_-}
\newcommand{\DAO}{\left[\partial \mathcal{A}\right]_0}
\newcommand{\A}{\mathcal{A}}

\begin{document}
\begin{frontmatter}

\title{On MPC without terminal conditions for dynamic non-holonomic robots}\thanks{This work has been submitted to IFAC for possible publication.}

\author[First]{Franz Ru{\ss}wurm} 
\author[First]{Willem Esterhuizen}
\author[Second]{Karl Worthmann}
\author[First]{Stefan Streif}

\address[First]{Technische Universität Chemnitz, Automatic Control and System Dynamics Laboratory, 
	Germany (e-mail: franz.russwurm, willem.esterhuizen, stefan.streif@etit.tu-chemnitz.de)}

\address[Second]{Technische Universität Ilmenau, Optimization-based Control, 
	Germany (e-mail: karl.worthmann@tu-ilmenau.de)}

\begin{abstract}                
	We consider an input-constrained differential-drive robot with actuator dynamics. For this system, we establish asymptotic stability of the origin on arbitrary compact, convex sets using Model Predictive Control (MPC) without stabilizing terminal conditions despite the presence of state constraints and actuator dynamics. We note that the problem without those two additional ingredients was essentially solved beforehand, despite the fact that the linearization is not stabilizable. We propose an approach successfully solving the task at hand by combining the theory of barriers to characterize the viability kernel and an MPC framework based on so-called cost controllability. 
	Moreover, we present a numerical case study to derive quantitative bounds on the required length of the prediction horizon. To this end, we investigate the boundary of the viability kernel and a neighbourhood of the origin, i.e.\ the most interesting areas.
\end{abstract}


\end{frontmatter}

\section{Introduction}

Differential-drive robots provide control problems of particular interest due to their non-holonomic constraints and varying complexity, ranging from basic two-dimensional examples to more complex ones that consider, e.g., acceleration, friction and constraints in form of a field of view \citep{Maniatopoulos2013}. The stabilization problem has been solved for such systems, even though the linearized system is often not stabilizable. Models of these nonholonomic systems are either kinematic, where inputs are velocities, or dynamic (and more complicated), where inputs are accelerations.

Various control schemes have been applied to such systems, including approaches that use probabilistic road maps and graphs \citep{Barraquand_1991,Kavraki_1996}, adaptive control \citep{Koubaa_2015} and neural networks \citep{Hu_2002}, as well as model predictive control (MPC) \citep{Gu2005, Gu2006, Xie2008,  Maniatopoulos2013, Worthmann2016, Bouzoualegh2018}.

MPC is an optimization-based control methodology, where the current state is measured, a finite-horizon optimal control problem is solved, and the first portion of the optimal solution is applied to the plant. The state is then measured again, and the process is iterated.
Two important aspects of MPC are \emph{asymptotic stability} of the closed loop and \emph{recursive feasibility} of the invoked finite-horizon optimization problems. A popular way to guarantee these two properties is to add a terminal cost and a control-invariant terminal set~\citep{Fontes_2001,RawlMayn18}. If the finite-horizon Optimal Control Problems (OCPs) are not augmented by such 
``stabilizing ingredients'', a sufficiently long prediction horizon is required in order to rigorously guarantee asymptotic stability and recursive feasibility invoking cost controllability, see, e.g.\ \citep{NeviPrim97,GrunPann10,BOCCIA2014} or \citep{REBLE2012,CoroGrun20,Esterhuizen2021} for extensions to continuous-time systems. 

The paper~\citep{Worthmann2016} applied continuous-time MPC without stabilizing ingredients to drive a three-dimensional non-holonomic robot to a desired position and orientation using a purely kinematic model. Hence, the robot was able to stop immediately 
such that recursive feasibility was trivially satisfied. 
In the current paper, we extend this study by considering a four-dimensional vehicle model, for which such a stopping input does not exist. Hence, we first establish local asymptotic stability using the framework proposed in~\citep{CoroGrun20} for homogeneous systems. Then, we argue that the globalization strategy presented in~\citep{Esterhuizen2021} is applicable and show that arbitrary compact, convex sets contained in the viability kernel can be rendered a domain of attraction, if the length of the prediction horizon is sufficiently long. To this end, an in-depth analysis of the viability kernel, i.e.\ the set of initial states for which there exists an input such that the constraints are satisfied for all future time, based on the theory of barries is needed.
Some other works that have investigated the use of the viability kernel, and related ideas, to the control of robotic vehicles include \citep{Bouguerra_2019, fraichard2004inevitable, kalisiak2004approximate,panagou2013viability}, but they do not study its role within MPC.

The outline of the paper is as follows. We give a brief summary of the theory of barriers in Section~\ref{sec:BarrierTheory}, which we apply on a model of a four-dimensional robot in Section~\ref{sec:viability} to precisly determine the viability kernel in view of a state constraint. In Section~\ref{sec:MPC_for_robot}, we ensure the local closed-loop stability of the robot near the origin under the control input provided by a nonlinear MPC algorithm, using recent results on cost controllability, a sufficient stability condition for systems with a controllable homogeneous approximation. Section~\ref{sec:Numerical_Results} then provides numerical results from simulations on the minimal stabilizing prediction horizon for the system starting from certain points of interest on the boundary of the viability kernel as well as using initial conditions near the origin. Conclusions are given in Section~\ref{sec:Conclusion}.
	
\section{Recap on the Theory of Barriers}\label{sec:BarrierTheory}

We briefly summarize the relevant theory from the paper \citep{DeDona_Levine} that we will use to characterize the viability kernel of the four-dimensional robot in Section~\ref{sec:viability}.
	
Consider the control system
\begin{align}
	\dot{x}(t) &= f(x(t),u(t)),	\,\, x(0) = x^0, \label{eq_NL_1}\\
	g_i(x(t)) &\leq 0 \; \; \forall \, t \in [0,\infty), \; i = 1,2,\ldots,q, \label{eq_NL_2}
\end{align}
with the state $x(t) \in \mathbb{R}^n$, the control $u(t) \in \mathbb{R}^m$ and $q$ state constraints $g_i: \mathbb{R}^n \rightarrow \mathbb{R}$. We assume that the control $u$ belongs to the set $\mathcal{U}$ of Lebesgue measurable functions that map the interval $[0, \infty)$ to a compact and convex set $\U \subset \mathbb{R}^m$. Moreover, we make the following assumptions as in \citep{DeDona_Levine}.
\begin{description}
\item[Assumptions]
\end{description}
\begin{itemize}
\item[(A1)] The function $f:\mathbb{R}^n\times\mathbb{R}^m\rightarrow\mathbb{R}^n$ is $\mathcal{C}^2$ on an open set containing $\mathbb{R}^n \times \mathbb{U}$.
\item[(A2)] There exists a $C<\infty$ such that $\sup_{u\in \U} |x^\top f(x,u)| \leq C (1 + \Vert x \Vert^2)$ for all $x\in\mathbb{R}^n$.
\item[(A3)] The set $f(x,\U)$ is convex for all $x\in\mathbb{R}^n$.
\item[(A4)] The function $g_i$ is $\mathcal{C}^2$ and the set of points given by $g_i(x) = 0$ defines an $n-1$ dimensional manifold.
\end{itemize}


With $x^{(u,x^0)}(t)$, or $x^{u}(t)$ if the initial state $x^0$ is clear from context, we denote the solution at time $t \geq 0$. To ease notation we introduce $\mathbb{I}(x) = \{ i \in \{1,2,\ldots,q\} \; \vert \; g_i(x) = 0 \}$, the set of active indices at $x \in \mathbb{R}^n$; $\mathbb{X} := \{ x \in \mathbb{R}^n \; \vert \; g_i(x) \leq 0 \; \; \forall \, i = 1,2,\ldots, q \}$, which denotes the \emph{feasible set}; $\mathbb{X}_0 := \{ x \in \mathbb{X} \; \vert \; \mathbb{I}(x)\neq\emptyset\}$ and $\mathbb{X}_- := \{ x \in \mathbb{X} \; \vert \; \mathbb{I}(x)=\emptyset\}$. The unit circle is denoted by $\mathbb{S}^1$. The notation $L_fg(x,u)$ denotes the Lie derivative of a differentiable function $g:\mathbb{R}^n\rightarrow \mathbb{R}$ at $x\in\mathbb{R}^n$ in the direction of $f(\cdot,u)$.

Next, we introduce the viability kernel, which is called the \emph{admissible set} in \citep{DeDona_Levine}.
\begin{definition}
	The \emph{viability kernel} is the set of initial states $x^0\in\mathbb{R}^n$, for which there exists an input $u\in\mathcal{U}$ such that the resulting integral curve satisfies the state constraints \eqref{eq_NL_2} for all future time, i.e.,
	\begin{equation*}
		\mathcal{A} := \{ x^0 \in \mathbb{X} \; \vert \; \exists \, u \in \mathcal{U}: \; x^{(u,x^0)}(t) \in \mathbb{X} \; \; \forall \, t \in [0, \infty) \}.
	\end{equation*}
\end{definition}
Under (A1)-(A4) the admissible set is closed. The main result from \citep{DeDona_Levine} is a characterization of the set's boundary, $\partial\A$, which consists of two complementary parts, $\DAO := \partial \A~\cap~\mathbb{X}_0$ and $\DAM:=\partial~\A~\cap \mathbb{X}_-$ called the \emph{usable part} and the \emph{barrier}, respectively. As shown in \citep[Th. 7.1]{DeDona_Levine}, the barrier is made up of integral curves of the system that satisfy a minimum-like principle.
\begin{theorem}\label{Main_Barrier_Theorem}
Under Assumptions (A1) - (A4) every integral curve $x^{\bar{u}}:[0,\infty)\rightarrow \mathbb{R}^n$ running along the barrier $\DAM$ and the corresponding control function $\bar{u}\in\mathcal{U}$ satisfy the following necessary conditions.
There exists a nonzero absolutely continuous maximal solution $\lambda^{\bar{u}}:[0,\infty)\rightarrow \mathbb{R}^n$ to the adjoint equation
\begin{equation}
	\dot{\lambda}^{\bar{u}}(t) = - \left( \frac{\partial f}{\partial x} (x^{\bar{u}}(t),\bar{u}(t)) \right)^T \lambda^{\bar{u}}(t),\label{eq:adjoint_equation}
\end{equation}
such that
\begin{align}
	\min_{u \in \U} \; \left\lbrace \lambda^{\bar{u}}(t)^T f(x^{\bar{u}}(t), u) \right\rbrace = \lambda^{\bar{u}}(t)^T f(x^{\bar{u}}(t), \bar{u}(t)) = 0,\label{eq:Hamiltonian}
\end{align}
for almost all $t\in[0,\infty)$. Moreover, if $x^{\bar{u}}$ intersects $\mathbb{X}_0$ we have $\lambda^{\bar{u}}(\bar{t}) = \nabla g_{i^*}(z)$, where
\begin{equation}
	\min_{u\in \U} \max_{i\in\mathbb{I}(z)} L_f g_i(z,u) = L_f g_{i^*}(z,\bar{u}(\bar{t}) )= 0,\label{cond_ult_tan}
\end{equation}
$\bar{t}<\infty$ denotes the time at which $\mathbb{X}_0$ is reached, and $z:= x^{\bar{u}}(\bar{t})$.
\end{theorem}
To provide some intuition, the barrier is that part of the viability kernel's boundary that is located in the interior of the feasible set, and is made up of ``extreme trajectories'' (that minimize the Hamiltonian a.e., as in \eqref{eq:Hamiltonian}), which may eventually intersect the boundary $\mathbb{X}_0$ in a tangential manner (as in \eqref{cond_ult_tan}). This part of the set's boundary is called the barrier because crossing it means that the state passes into the complement of $\A$, from where constraint violation is unavoidable regardless of the future control.

One can describe a system's barrier using Theorem~\ref{Main_Barrier_Theorem} according to the following steps. Identify the points of ultimate tangentiality with condition \eqref{cond_ult_tan}, then integrate the system dynamics \eqref{eq_NL_1} and adjoint equation \eqref{eq:adjoint_equation} \emph{backwards in time} with the control function that minimizes the Hamiltonian for almost every $t$, as in condition \eqref{eq:Hamiltonian}.

\section{Viability kernel for a 4-D Nonholonomic Vehicle}\label{sec:viability} 

Consider the four-dimensional robot model given by \eqref{eq_NL_1} with
\begin{equation}\label{eq:4D_Robot}
	f(x,u) = 
	\left(
	\begin{array}{c}
		x_4(t) \cos(x_3(t))\\
		x_4(t) \sin(x_3(t))\\
		u_1(t)\\
		u_2(t)
	\end{array}
	\right)	
\end{equation}
where $(x_1,x_2)\in\mathbb{R}^2$ is the car's location, $x_3\in\mathbb{S}^1$ is the car's orientation and $x_4\in\mathbb{R}$ is the car's speed. The input must satisfy $u(t)\in \U$  for all $t\geq 0$, where
\begin{equation}
\U:= \{ u \in \mathbb{R}^2 \; \vert \; \underline{u}_i \leq u_i(t) \leq \overline{u}_i \, , \, i = 1,2 \},\label{eq:InputConstraints} 
\end{equation}
$\underline{u}_i\in\mathbb{R}_{< 0}$ and $\overline{u}_i\in\mathbb{R}_{> 0}$. Thus, $\overline{u}_1$ (resp. $\underline{u}_1$) is the maximal (resp. minimal) rate of clockwise rotation, and $\overline{u}_2$ (resp. $\underline{u}_2$) is the maximal (resp. minimal) acceleration. We impose the state constraint:
\begin{equation}
	g(x(t)) = x_1(t) - 1 \qquad\forall\, t\geq 0.\label{wall_constraint}
\end{equation}

We remark that this is an extension of the classic three-dimensional Dubins' vehicle, which can be obtained if the speed, 
i.e.\ the $x_4$-component, is kept constant. %
Similarly, it may be considered as an extension of the non-holonomic robot, if $x_4$ is considered as the second control. %
This corresponds to a purely kinematic model of a mobile robot, cp.\ \citep{Worthmann2016}.

We observe that the set $\{x\in\mathbb{R}^4: x_1\leq 1, x_4 = 0\}$ is the set of controlled equilibria (using $u_1 = u_2 \equiv  0\in \U$) meaning, in particular, that it can be rendered invariant. Thus, the interesting phenomena appear when $x_4\neq 0$. Invoking ultimate tangentiality~\eqref{cond_ult_tan}, we get:
\begin{equation*}
	\min_{u \in \mathbb{U}} \; L_fg(z,u) = z_4 \cos(z_3) \stackrel{!}{=} 0,
\end{equation*}
c.p. \eqref{cond_ult_tan}. Thus, supposing $z_4 \neq 0$, we have $z_3 = \pm \frac{\pi}{2}$, giving points of ultimate tangentiality $z = (1, z_2, \pm\frac{\pi}{2}, z_4)^T$ with $z_2\in \mathbb{R}, z_4\in \mathbb{R}\setminus\{0\}$. This fits intuition: if the car arrives at the wall with nonzero speed (along an ``extremal'' trajectory), then it must just brush past it to avoid collision. It will be convenient to refer to the following four sets of tangent points associated with a nonzero final speed. 
\begin{align*}
	\mathcal{T}_1 &:=\{z\in\mathbb{R}^4 \vert z_1=1, z_3 = \frac{\pi}{2}, z_4 >0\},\\ 
	\mathcal{T}_2 &:=\{z\in\mathbb{R}^4 \vert z_1=1, z_3 = \frac{\pi}{2}, z_4 <0\},\\ 
	\mathcal{T}_3 &:=\{z\in\mathbb{R}^4 \vert z_1=1, z_3 = -\frac{\pi}{2}, z_4 >0\},\\ 
	\mathcal{T}_4 &:=\{z\in\mathbb{R}^4 \vert z_1=1, z_3 = -\frac{\pi}{2}, z_4 <0\}.
\end{align*}

The control input for barrier trajectories is obtained from \eqref{eq:Hamiltonian}. Considering
\begin{equation*}
\min_{u \in \mathbb{U}} \; \left(\lambda_3(t) u_1 + \lambda_4(t) u_2\right),
\end{equation*}
we get:
\begin{align}\label{example_control-laws}
\begin{split}
u_1(t) = \begin{cases} 
\overline{u}_1 &\mbox{if } \lambda_3(t)<0,\\
\underline{u}_1 & \mbox{if } \lambda_3(t)>0,\\
\mbox{anything} & \mbox{otherwise},
\end{cases} 
\\
u_2(t) = \begin{cases} 
\overline{u}_2 &\mbox{if } \lambda_4(t)<0,\\
\underline{u}_2 & \mbox{if } \lambda_4(t)>0,\\
\mbox{anything} & \mbox{otherwise}.
\end{cases}
\end{split}
\end{align}
Thus, along the barrier trajectories full acceleration or breaking as well as full left or right steering is applied.
The adjoint equation \eqref{eq:adjoint_equation} reads
\begin{equation*}
	\dot{\lambda}(t) = 
	\left(
	\begin{array}{c}
		0\\
		0\\
		x_4(t)\sin(x_3(t))\lambda_1(t) -x_4(t)\cos(x_3(t))\lambda_2(t)\\
		-\cos(x_3(t))\lambda_1(t) - \sin(x_3(t))\lambda_2(t)
	\end{array}
	\right)	
\end{equation*}
with $\lambda(\bar{t}) = \nabla g(z) = (1,0,0,0)^T$. With this information, and the fact that analytic solutions to the dynamics are easily obtainable for constant inputs, we arrive at the following result.
\begin{proposition}\label{prop_1}
	Every curve running along the barrier, $\DAM$, that ends at a tangent point contained on:
	\begin{itemize}
		\item  $\mathcal{T}_1$ is given by $u_1(t) = \overline{u}_1$, $u_2(t) = \underline{u}_2$,
		\item $\mathcal{T}_2$ is given by $u_1(t) = \overline{u}_1$, $u_2(t) = \overline{u}_2$,
		\item $\mathcal{T}_3$ is given by $u_1(t) = \underline{u}_1$, $u_2(t) = \underline{u}_2$,
		\item $\mathcal{T}_4$ is given by $u_1(t) = \underline{u}_1$, $u_2(t) = \overline{u}_2$,
	\end{itemize}
for $t\in[\hat{t},\bar{t}]$, $\hat{t} := \bar{t} - \frac{\pi}{\overline{u}_1 - \underline{u}_1}$. Moreover, at $\hat{t}$ every barrier curve ending on $\mathcal{T}_1$ intersects a barrier curve ending on $\mathcal{T}_3$, and every barrier curve ending on $\mathcal{T}_2$ intersects a barrier curve ending on $\mathcal{T}_4$.
\end{proposition} 

\begin{figure}[ht]
	\begin{center}
		\includegraphics[width=7.2cm]{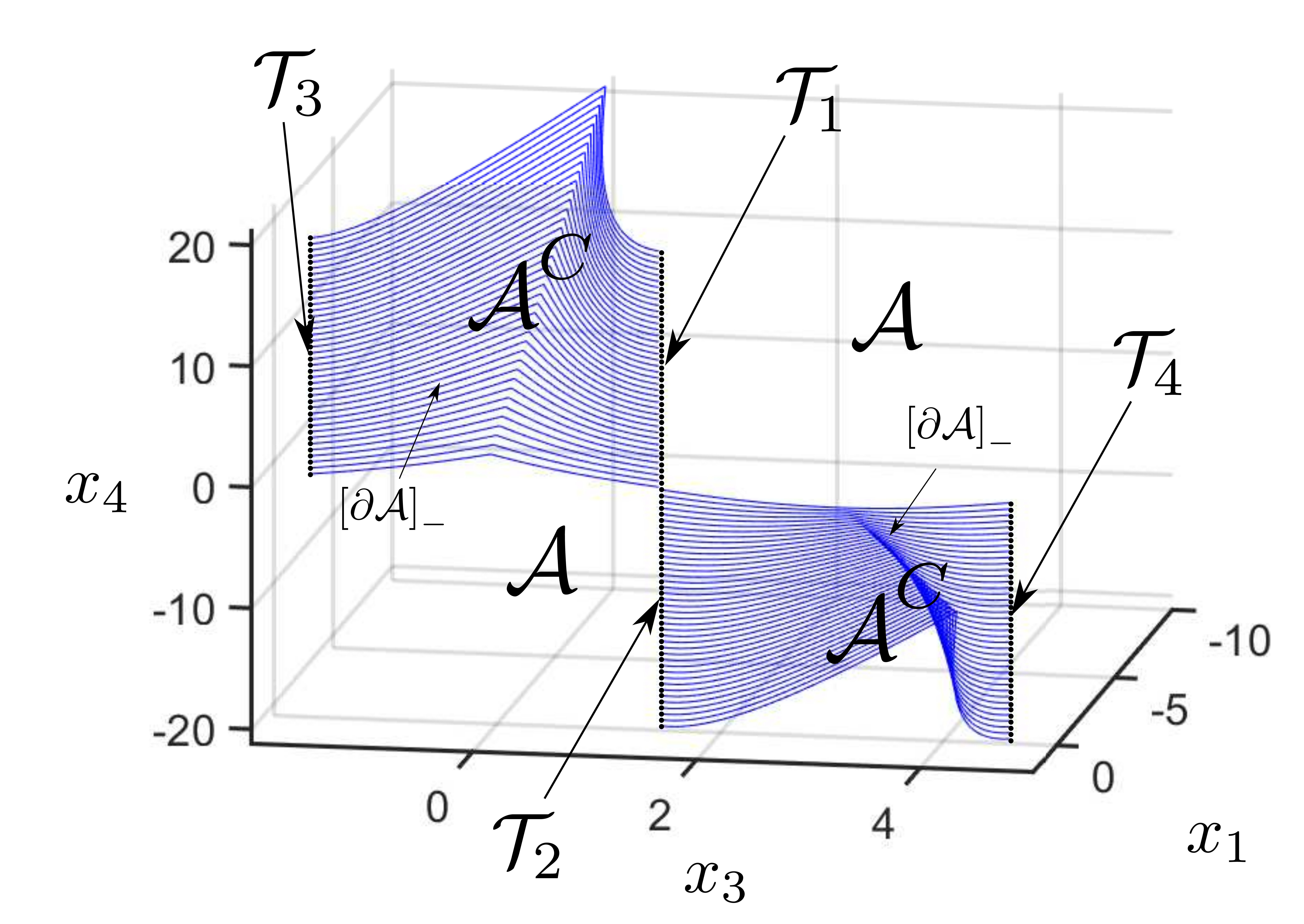}    
		\caption{Viability kernel of system \eqref{eq:4D_Robot}, with constraints \eqref{eq:InputConstraints} and \eqref{wall_constraint} described by curves satisfying Theorem~\ref{Main_Barrier_Theorem}.}\label{fig:Admissible_Set_Robot}
	\end{center}
\end{figure}

Figure~\ref{fig:Admissible_Set_Robot} shows the viability kernel for the vehicle with $-\underline{u}_i = \overline{u}_i = 2$, $i=1,2$. We sample final points from the four line segments of tangent points, $\mathcal{T}_1$, $\mathcal{T}_2$, $\mathcal{T}_3$ and $\mathcal{T}_4$, and integrate the system \emph{backwards in time} utilizing the input as in Proposition~\ref{prop_1} for $t\in[-\frac{\pi}{4},0]$. At $t=-\frac{\pi}{4}$ the curves intersect at \emph{stopping points}, \citep{esterhuizen2014_stopping_pts}, which form the kink in the figure. Further backward integration of the curves from these points need to be ignored, because they pass into the interior of the kernel. From this non-differentiable part of the barrier there are two extremal admissible inputs: one that causes the car to arrive at the wall with orientation $x_3(\bar{t})=\frac{\pi}{2}$, and the other that causes it to arrive with orientation $x_3(\bar{t}) = -\frac{\pi}{2}$.
\begin{remark}\label{rem:constraints:multiple}
	The single state constraint \eqref{wall_constraint} may be generalized to the constraint $a_1 x_1(t) + a_2 x_2(t) \leq b$, $\forall t \geq 0$, with coefficients $a_1, a_2, b \in \mathbb{R}$ if a suitable transformation is applied that rotates and translates the coordinates $x_1$ and $x_2$ appropriately. In the future we intend to extend the analysis to multiple planar constraints. We believe that the obtained barrier curves intersect more elaborately in this case.
\end{remark}
\begin{remark}\label{rem:stopping}
	The following stop maneuver will be needed in the next section to prove stability of the MPC closed-loop, i.e., we show the system is finite-time null controllable on the viability kernel. Initiating at an initial state $x(0)\in\A$ the car can be made to stop (that is, reach an arbitrary state with component $x_4 = 0$) in finite time. If $x_4(0) > 0$, employ $u_2(t) = \underline{u}_2$. If the state reaches the barrier, let $u_1(t)$ switch to the appropriate full left/right turn depending on which part of the barrier is reached. The car will then reach the wall in finite time and pass into $\A$'s interior. Keep employing this switching law (the state may reach the barrier multiple times) until $x_4 = 0$, then let $u_2 = 0$. Use an analogous maneuver for $x_4(0) < 0$.
\end{remark}
\section{MPC: closed-loop stability}\label{sec:MPC_for_robot}

In this section, we show that system \eqref{eq:4D_Robot} is asymptotically stable under the MPC closed loop on compact and convex sets containing the origin, i.e.\ a (controlled) equilibrium in the interior of the viability kernel, for suitably designed stage cost provided that the prediction horizon is sufficiently long (without using terminal ingredients in the optimization problem). The restriction to bounded sets is mainly needed to exclude pathological cases, e.g.\ arbitrarily large initial speed, since we do not employ the information on the viability kernel deduced in the previous section in the MPC algorithm. 
We apply the framework proposed in~\citep{CoroGrun20} to rigorously show asymptotic stability of the origin using the homogeneous approximation at the origin. Moreover, we use the globalization strategy introduced in~\citep{Esterhuizen2021} to enlarge the domain of attraction. 

Consider the finite-horizon cost functional:
\[
	J(\hat{x},u) = \int_{0}^T \ell(x^{(u,\hat{x})}(s),u(s))\,\mathrm{d}s,
\]
where $\ell:\mathbb{R}^n\times\mathbb{R}^m\rightarrow\mathbb{R}$ is the continuous stage cost. Let $\mathcal{U}_T(\hat{x})$ denote the set of all $u\in\mathcal{U}$ such that the solution to \eqref{eq:4D_Robot} satisfies the constraint \eqref{wall_constraint} for all $t\in[0,T]$. The MPC algorithm reads as follows,
\begin{alg}[MPC]\label{Alg:MPC}\ \\
	\textbf{Input}: time shift $\delta \in \mathbb{R}_{>0}$, $N \in \mathbb{N}$, initial state $x^0 \in \mathcal{A}$\\
	\textbf{Set}: prediction horizon $T \leftarrow N \delta$, $\hat{x} \leftarrow x^0$.
	\begin{enumerate}
		\item Find a minimizer $u^\star_{\hat{x}} \in \arginf_{u \in \mathcal{U}_T(\hat{x})} J(\hat{x},u)$
		\item Implement $u^\star_{\hat{x}}(t)$, for $t \in [0, \delta)$
		\item Set $\hat{x}\leftarrow x^{(u^\star,\hat{x})}(\delta)$ and go to step (1)
	\end{enumerate}
\end{alg}
With slight abuse of notation, in Step (3) we let $x^{(u^\star,\hat{x})}(t)$ denote the solution at $t$ from $\hat{x}$ with $u^{\star}_{\hat{x}}$. For given prediction horizon~$T$ and initial value~$\hat{x}$, 
the infimum of the \textit{parametric} optimal control problem to be solved in Step~(1) of Algorithm~\ref{Alg:MPC} is denoted by~$V_T(\hat{x})$, %
which allows us to implicitly define the (optimal) value function $V: \mathbb{X} \rightarrow \mathbb{R} \cup \{\pm \infty\}$. We assume that, if an infimum of the optimal control problem in Step~(1) exists, then it is attained by an admissible control, 
see \cite[p.56]{GrunPann17} for a discussion on this issue.

The MPC algorithm produces a sampled-data feedback law $\mu_{T,\delta}(t,x) = u^{\star}_{x}(t)$, $t\in[0,\delta)$. The resulting solution initiating from $x^0$ due to this feedback is denoted $x_{\mu_{T,\delta}}^{x^0}$.


We use \citep[Theorem~4.4]{CoroGrun20} to show asymptotic stability of the origin w.r.t.\ the MPC closed-loop. We refer to \citep[Sections~2 and~4]{CoroGrun20} for the definition of homogeneity and homogeneous approximation.
\begin{theorem}\label{Th:origin_asymp_stable}
	Consider a vectorfield~$f$ with homogeneous approximation~$h$ of degree $\tau \leq 0$. %
	Assume that the homogeneous system with vectorfield~$h$ is globally asymptotically null controllable to the origin. %
	Then there exists a neighborhood~$\mathcal{N}$ of the origin such that cost controllability holds, i.e.\ %
	there exists a monotonically increasing, bounded function $B: \mathbb{R}_{\geq 0} \rightarrow \mathbb{R}_{\geq 0}$ such that
	\begin{equation}\label{eq:CostControllability}
		V_{\bar{T}}(x^0) \leq B(\bar{T}) \cdot \ell^\star(x^0) \qquad\forall\,x \in \mathcal{N}, \bar{T} \geq 0
	\end{equation}	
	with $\ell^\star(x^0) := \min_{u \in \mathbb{R}^2} \ell(x^0,u)$. In particular, for given $\delta > 0$, %
	the origin is asymptotically stable w.r.t.\ the MPC closed-loop for a sufficiently large prediction horizon~$T$.
\end{theorem}

As a preliminary step, we show that the origin is \textit{globally} asymptotically stable for the homogeneous approximation of the System~\eqref{eq:4D_Robot}, i.e.
\begin{align}\nonumber
	\dot{x}(t) & = \begin{pmatrix}
		x_4(t) \\ x_3(t) x_4(t) \\ u_1(t) \\ u_2(t)
	\end{pmatrix} = f(x(t)) + \sum_{i=1}^2 g_i(x(t)) u_i(t) \\ 
	& := \begin{pmatrix}
		x_4(t) \\ x_3(t) x_4(t) \\0 \\ 0
	\end{pmatrix}  + \begin{pmatrix}
		0 \\ 0 \\ 1 \\ 0
	\end{pmatrix} u_1(t) + \begin{pmatrix}
		0 \\ 0 \\ 0 \\ 1
	\end{pmatrix} u_2(t), \label{eq:HomogeneousApproximation}
\end{align} 
if state and control constraints are ignored. 
\begin{remark}
	Note that neither the original dynamics~\eqref{eq:4D_Robot} nor the dynamics governed by the homogeneous approximation satisfy Brockett's condition %
	meaning that there does not exist a stabilizing, continuous static-state feedback law. %
	Moreover, note that already the simplified three-dimensional kinematic version of the example, where $x_4(t)$ is replaced by~$u_2(t)$, is not stabilizable %
	using MPC based on purely quadratic stage cost as rigorously shown in \cite[Subsection~4.1]{MullWort17}. %
	The same holds true for the respective homogeneous approximation, see~\cite[Proposition~2.5]{CoroGrun20}.
\end{remark}

We establish null controllability of System~\eqref{eq:HomogeneousApproximation}. %
W.l.o.g.\ let us suppose that $x_4^0 \neq 0$ holds. %
Otherwise, an infinitesimal small impulse via $u_2$ ensures this assumption. %
By setting $u_2 \equiv 0$, we ensure stationarity of the $x_4$-component. %
Hence, the $x_2$-$x_3$-system is essentially the double integrator, which is a linear controllable system. %
This implies, for arbitrary~$x^0 \in \mathbb{X}$, the existence of a time $\bar{t} \in [0,\infty)$ and a control function~$u_1:[0,\bar{t}]$ such that $x_2(\bar{t}) = x_3(\bar{t}) = 0$ holds, see, e.g.\ \cite[Chapter~2]{MackStra12}. %
Indeed, this control function is piece-wise continuous with finitely many switches. %
Then, we repeat this line of reasoning for the $x_1$-$x_4$-system setting $u_1 \equiv 0$ to ensure that the $x_2-$ and $x_3$-component remain at zero. In conclusion, the homogeneous approximation is globally null controllable.

Note that System~\eqref{eq:HomogeneousApproximation} is $(r,s,\tau)$-homogeneous with degree of homogeneity $\tau = 0$, $r_1=r_3=r_4=1=s_1=s_2=1$, and $r_2=2$ according to \cite[Definition~2.3]{CoroGrun20}. In conclusion, all assumptions of \cite[Theorem~3.5]{CoroGrun20} hold for the (compatible) stage cost
\begin{equation}\label{eq:StageCost}
	\ell(x,u) = x_1^4 + x_2^2 + x_3^4 + x_4^4 + u_1^4 + u_2^4,
\end{equation}
which allows us to conclude global asymptotic stability of the origin of \eqref{eq:HomogeneousApproximation} w.r.t.\ the MPC closed loop provided that the prediction horizon is sufficiently long. Note that the powers used in~\eqref{eq:StageCost} may be scaled with an arbitrary positive constant, e.g.\ penalizing all terms quadratically except for a linear penalization of the deviation w.r.t.\ the $x_2$-coordinate works as well. 

Finally, verifying that the homogeneous system~\eqref{eq:HomogeneousApproximation} is indeed the homogeneous approximation, allows us to infer \textit{local} asymptotic stability for System~\eqref{eq:4D_Robot} by applying Theorem~\ref{Th:origin_asymp_stable} noting that $0 \in \operatorname{int}(\mathbb{U})$ and that the origin is also contained in the interior of the viability kernel,~$\mathcal{A}$. This verification can be done analogously to the three-dimensional kinematic version of the example, see ~\cite[Proposition~4.6]{CoroGrun20} for details. 

To extend the local asymptotic stability to any bounded subset of the viability kernel, the globalization strategy proposed in~\cite{BOCCIA2014} and extended to continuous-time systems in~\citet{Esterhuizen2021} can be applied in order to enlarge the domain of attraction by showing cost controllability on arbitrary but fixed compact sets, $K \subset \mathcal{A}$. %
Since stopping ($x_4 = 0$) is possible for each $x^0 \in \mathcal{A}$ as explained in Remark~\ref{rem:stopping}, the following maneuver yields a uniform bound for the infinite-horizon value function~$V_\infty$ on~$K$, which is also an upper bound for~$V_{\bar{T}}$, $\bar{T}>0$: %
Stop, turn towards the desired equilibrium (the origin) using~$u_1$ (and $u_2 \equiv 0$), drive and stop at the origin using~$u_2$ (and $u_1 \equiv 0$), and turn again using~$u_1$ ($u_2 \equiv 0$). Since the stage cost is also lower bounded outside the neighborhood of the origin, on which we already proved asymptotic stability w.r.t.\ the MPC closed loop, the arguments used in~\citep{Esterhuizen2021} are directly applicable for this \textit{globalization strategy}. Essentially, this mimics the argumentation used in~\citep{Worthmann2016} in detail but extended by the stopping maneuver. The necessity to include this \textit{stopping} is also the reason for restricting ourselves to bounded sets to simplify the analysis.


\section{Numerical Results}\label{sec:Numerical_Results}

	In this section the NMPC routine from \citep{GrunPann17} is applied to the system \eqref{eq:4D_Robot} with the stage costs \eqref{eq:StageCost} to determine a stabilizing prediction horizon for certain points of interest. We implement the MPC algorithm on the discretized dynamics obtained with a zero-order-hold input with sampling time $0.02$s.
	
	For the first simulation, the stabilizing prediction horizon $\hat{N}$ for points on the boundary in the neighborhood of a non-differentiable kink of the viability kernel, where two barrier trajectories intersect, is investigated.
	The point $x^0$ is determined by utilizing the viability kernel calculated in Section~\ref{sec:viability}, with $\mathbb{U}=\{u \in \mathbb{R}^2 \; \vert \; u_i \in [-2,2] , \; i = 1,2 \}$. The results can be found in Table~\ref{tb:minimal_time_horizons_near_kink}, suggesting the horizon length $\hat{N}$ increases near non-differentiable parts of the kernel's boundary.
	
\begin{figure}[ht]
	\begin{center}
	\includegraphics[width=7.2cm]{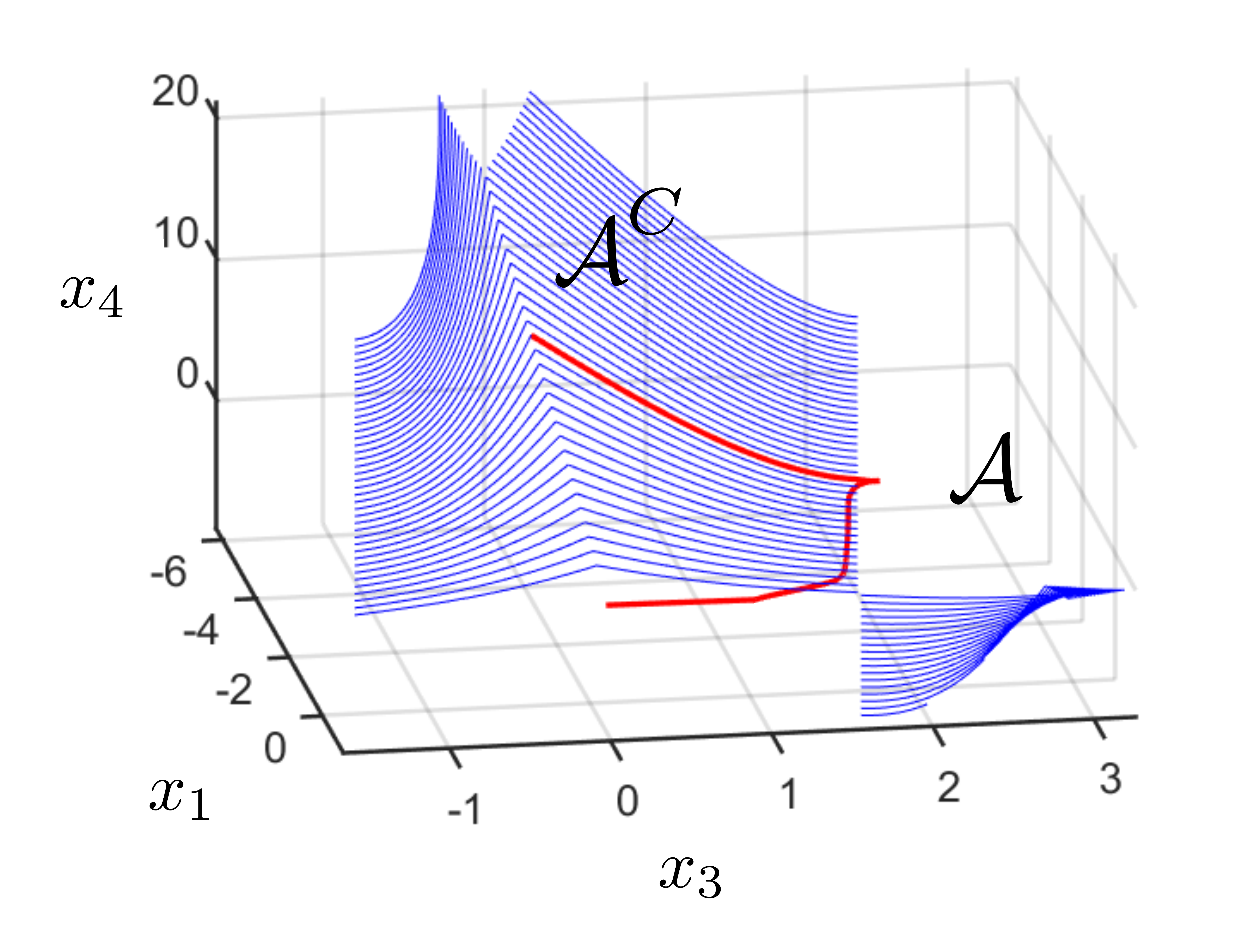}    
	\caption{Trajectory (red) of the system \eqref{eq:4D_Robot} generated by MPC starting in a kink on the boundary of the viability kernel $\mathcal{A}$ (blue).} 
	\label{fig:Trajectory_on_boundary}
	\end{center}
\end{figure}

	In the second simulation the initial conditions $x^0$ are located on the boundary of the viability kernel in the non-differentiable kinks. The results are shown in Table~\ref{tb:minimal_time_horizons_in_kink} and suggest that there might not be an upper bound for $\hat{N}$ on $\A$. Moreover, the stabilizing horizon increases with larger initial $x_4$ (speed) and more negative initial $x_1$ (initial distance from the wall). A trajectory of the nonholonomic vehicle, generated by MPC, starting at the point $x^0 = (-3.55, 0, 0, 9.67)^\top$ in a kink on the boundary of the viability kernel is shown in Figure \ref{fig:Trajectory_on_boundary}.
	The closed-loop solution runs along the boundary following exactly the barrier trajectory with the extremal input $u = (2,-2)^\top$.
	Note that $x^0_2$, in the simulations chosen as $x^0_2=0$, can be arbitrary in both simulations and does not effect the viability kernel nor viable trajectories, other than a parallel displacement along the $x_2$-axis.
	
	The third simulation, with the results presented in Table~\ref{tb:minimal_time_horizons_approaching_boundary}, examines the stabilizing prediction horizon of a series of points approaching the boundary at $x^0 = (-3.55, 0, 0, 9.67)^\top$ from inside the viability kernel.
	The prediction horizon $\hat{N}$ increases as the initial conditions come close to the boundary, supporting the assumption that the longest stabilizing prediction horizon can be expected on the boundary of the viability kernel.
	Here only $x^0_1$ is varied, providing the system more space to act before it eventually intersects with the constraints.
	
	The last simulation investigates the minimal prediction horizon for initial values approaching the origin for a sampling time of $1$s.
	We compare the horizon $\hat{N}$ using the stage costs \eqref{eq:StageCost} with the horizon $\tilde{N}$ utilizing the purely quadratic stage costs,	
	\begin{align}\label{eq:QuadraticStageCost}
	\ell(x,u) = x_1^2 + x_2^2 + x_3^2 + x_4^2 + u_1^2 + u_2^2.
	\end{align}	
	The results in Table~\ref{tb:minimal_time_horizons_near_origin} show a remarkably smaller stabilizing prediction horizon for the stage costs \eqref{eq:StageCost}, justifying the choice of these costs over the quadratic costs \eqref{eq:QuadraticStageCost}, see \citet{Worthmann2016}.
	
	\begin{table}[hb]
		\begin{center}
		\caption{Stabilizing prediction horizon~$\hat{N}$ for initial values near a kink, $x^0_2~=~0$}\label{tb:minimal_time_horizons_near_kink}
		\begin{tabular}{c|rrrrr}
			$x^0_1$ & $-1.878$ & $-3.261$ & $-3.55$ & $-3.261$ & $-1.878$ \\\hline
			$x^0_3$ & $0.36$ & $0.06$ & $0$ & $-0.06$ & $-0.36$ \\\hline
			$x^0_4$ & $9.31$ & $9.61$ & $9.67$ & $9.61$ & $9.31$ \\\hline
			$\hat{N}$ & $25$ & $32$ & $34$ & $32$ & $25$
			
		\end{tabular}
		\end{center}
	\end{table}
	
	\begin{table}[hb]
		\begin{center}
		\caption{$\hat{N}$ for initial values in a kink, $x^0_2~=~x^0_3~=0$}\label{tb:minimal_time_horizons_in_kink}
		\begin{tabular}{c|rrrrr}
			$x^0_1$ & $-0.05$ & $-1.8$ & $-3.55$ & $-5.05$ & $-6.05$ \\\hline
			$x^0_4$ & $2.67$ & $6.17$ & $9.67$ & $12.67$ & $14.67$ \\\hline
			$\hat{N}$ & $30$ & $33$ & $34$ & $35$ & $36$
			
		\end{tabular}
		\end{center}
	\end{table}
	
	\begin{table}[hb]
		\begin{center}
		\caption{$\hat{N}$ for initial values approaching boundary with $x^0~=~(x^0_1, 0, 0, 9.67)^\top$}\label{tb:minimal_time_horizons_approaching_boundary}
		\begin{tabular}{c|rrrrr}
			$x^0_1$ & $-3.85$ & $-3.7$ & $-3.65$ & $-3.6$ & $-3.55$ \\\hline
			$\hat{N}$ & $29$ & $29$ & $30$ & $31$ & $34$
			
		\end{tabular}
		\end{center}
	\end{table}
	
	\begin{table}[hb]
		\begin{center}
		\caption{$\hat{N}$ with stage costs \eqref{eq:StageCost} and $\tilde{N}$ with stage costs \eqref{eq:QuadraticStageCost} for initial values $x^0~=~(0, x^0_2, 0, 0)^\top$}\label{tb:minimal_time_horizons_near_origin}
		\begin{tabular}{c|cccccccc}
			$x^0_2$ & $2$ & $1$ & $2^{-1}$ & $2^{-2}$ & $2^{-3}$ & $2^{-4}$ & $2^{-5}$ & $2^{-6}$ \\\hline
			$\hat{N}$ & $5$ & $6$ & $6$ & $6$ & $6$ & $7$ & $7$ & $7$ \\
			$\tilde{N}$ & $19$ & $27$ & $31$ & $31$ & $32$ & $33$ & $34$ & $34$	
		\end{tabular}
		\end{center}
	\end{table}
	
\section{Conclusion}\label{sec:Conclusion}
	
We applied the theory of barriers to determine the viability kernel of a four-dimensional mobile robot and ensured asymptotic stability w.r.t.\ the MPC closed loop of the viability kernel on arbitrary compact and convex sets containing the origin 
based on results from~\citep{CoroGrun20} and~\citep{Esterhuizen2021}. Furthermore, numerical tests on the minimal stabilizing prediction horizon were carried out for certain points of interest of the viability kernel.

The results suggest that, for systems with unbounded viability kernels, there might not exist an upper bound for the \textit{sufficiently long} prediction horizon in MPC without terminal conditions. In the future, we intend to incorporate knowledge of the viability kernel to shorten this horizon. In particular, we intend to investigate the effects of replacing the state constraints $\mathbb{X}$ with the kernel $\mathcal{A}$ in the MPC algorithm.

\begin{ack}
The project underlying this paper was funded by the Federal Ministry of Education and Research under the grant number 281B300116. Karl Worthmann gratefully acknowledges funding by the German Research Foundation (DFG; grant WO 2056/6-1, project number 406141926)
\end{ack}

\bibliography{bib/ifacconf_1}             

\end{document}